\documentclass[conference]{IEEEtran}
\IEEEoverridecommandlockouts
\renewcommand\IEEEkeywordsname{Keywords}
\ifCLASSINFOpdf
\else
\fi
\usepackage{xcolor,soul,framed} 
\colorlet{shadecolor}{yellow}
\usepackage[pdftex]{graphicx}
\graphicspath{{../pdf/}{../jpeg/}}
\DeclareGraphicsExtensions{.pdf,.jpeg,.png}
\usepackage{cite}
\usepackage{amsmath,amssymb,amsfonts}
\usepackage{algpseudocode}
\PassOptionsToPackage{ruled, linesnumbered}{algorithm2e}
\usepackage{algorithm2e}
\SetKw{Break}{break out of the loop}
\usepackage{graphicx}
\usepackage{textcomp}
\usepackage{comment}
\usepackage{units}
\usepackage{url}
\usepackage{geometry}
\usepackage{xcolor}
\definecolor{myblue}{rgb}{0.0, 0.5, 1.0}
\definecolor{myred}{rgb}{1.0, 0.13, 0.32}
\definecolor{mygreen}{rgb}{0.31, 0.68, 0.07}
\usepackage[pdftex]{graphicx}
\DeclareGraphicsExtensions{.pdf,.jpeg,.png}
\def\BibTeX{{\rm B\kern-.05em{\sc i\kern-.025em b}\kern-.08em
    T\kern-.1667em\lower.7ex\hbox{E}\kern-.125emX}}
    \DeclareMathOperator*{\argmax}{arg\,max}  

\usepackage{pgfplots}
\usepackage{pgfplotstable}
\usepackage{tikz}
\usetikzlibrary{calc}

\usepackage{balance}

\begin{document}

\title{ {\huge On Energy Efficiency and Fairness Maximization in
RIS-Assisted MU-MISO mmWave Communications }
\thanks{This publication has emanated from research conducted with the financial support of Science Foundation Ireland under Grant Number 13/RC/2077\_P2.}
}
\newgeometry {top=25.4mm,left=19.1mm, right= 19.1mm,bottom =19.1mm}
\author{Ahmed Magbool, Vaibhav Kumar, and Mark F. Flanagan \\
School of Electrical and Electronic Engineering, University College Dublin, Belfield, Dublin 4, Ireland \\
Email: ahmed.magbool@ucdconnect.ie, vaibhav.kumar@ieee.org, mark.flanagan@ieee.org}
\maketitle
\begin{abstract}
Reconfigurable intelligent surfaces (RISs) are considered to be a promising solution to overcome the blockage issue in the millimeter-wave (mmWave) band. Energy efficiency is an important performance metric in mmWave systems with a large number of antennas. However, due to the severe path loss in mmWave systems, resource allocation algorithms tend to allocate most of the resources for the benefit of the users with higher channel gains. In this paper, we propose a lexicographic-based approach to find the optimal power allocation, RIS phase shift matrix, and analog precoders that maximize both energy efficiency and user fairness. We solve the corresponding multi-objective optimization problem in two stages. In the first stage, we maximize the energy efficiency, and in the second stage we maximize the fairness subject to a minimum energy efficiency constraint. We propose an alternating optimization procedure to solve the optimization problem in each stage. The optimal power allocation is found using Dinkelbach's method in the first stage and using convex optimization techniques in the second stage, the RIS phase shift matrix is found using a gradient ascent algorithm, and the analog precoder is determined using beam alignment. Numerical results show that the proposed algorithm can achieve an excellent trade-off between the energy efficiency and fairness by boosting the minimum weighted rate with a minor and controllable reduction in the energy efficiency. 
\end{abstract}
\IEEEpeerreviewmaketitle
\section{Introduction}
 To cope with the ever-increasing demand for better connectivity and higher data rates, the millimeter-wave (mmWave) band has been investigated as a spectrum enabler for the existing fifth-generation (5G) and the upcoming sixth-generation (6G) of cellular wireless communication networks \cite{6G_survey}. To address the high sensitivity to blockages in the mmWave band, the use of reconfigurable intelligent surfaces (RISs) has been proposed to create a programmable propagation environment by controlling the phase of the incident signals using low-cost nearly-passive reflecting elements~\cite{ RIS_mot3}.
  
Recent research contributions draw more attention to the energy efficiency of mmWave systems \cite{mmwave_comm_survey}, where the spectral efficiency should be enhanced at a minimal expense in additional power consumption. Two solutions for EE maximization are proposed in \cite{EE_opt3}; one is gradient-based and the other uses a  sequential fractional programming based alternating approach. A genetic approach is proposed in \cite{EE_opt4} based on a covariance matrix adaptation evolution strategy and Dinkelbach's method to maximize the EE in mmWave and terahertz (THz) systems. The optimal beamformer design is obtained in \cite{EE_opt5} along with an asymptotic channel capacity characterization under hardware impairments. Furthermore, a long short-term memory (LSTM) deep neural network has been utilized to boost the EE in \cite{EE_opt6}.
 
 Due to the high path loss in mmWave systems, users served by the same base station (BS) can have significantly different channel strengths. When user fairness is ignored, EE optimization algorithms tend to give users with very small channel gains a very low quality of service (QoS), as these users make only a minor contribution to the EE objective function. One solution to avoid this is to incorporate QoS constraints into the optimization problem, e.g., by forcing the rate of each user to be greater than a specific threshold. However, determining these threshold rates can be problematic as it can make the optimization problem infeasible. Moreover, the simulation results in \cite{EE_opt3} show that setting high thresholds can severely impact the EE of the system. To avoid this, a multi-objective function is formulated in~\cite{fairness_1} which combines the sum rate and Jain's fairness index; a successive convex approximation method is proposed to solve the resulting non-convex optimization problem. The minimum rate in the system is maximized in~\cite{fairness_2} whereas the minimum EE of the individual users is targeted in~\cite{fairness_3}.
 
 In this paper, we target the maximization of both EE and fairness in RIS-assisted multi-user multiple-input-single-output (MU-MISO) mmWave systems. First, we formulate a multi-objective optimization which focuses on maximizing the system's EE as well as the fairness between users, which is expressed in terms of proportional rates. The former is an important aspect of system design while the latter is crucial to assuring the QoS of end users. Next, we use a lexicographic method to divide the optimization problem into two stages. In the first stage, we maximize the EE only, and in the second stage, we maximize the fairness subject to an EE constraint derived from the solution found in the first stage. In each stage, we use an alternating optimization (AO) framework to find the optimal power allocation, RIS phase shift matrix and analog precoders. We provide extensive numerical results to evaluate the performance of the proposed method, and demonstrate its improved performance with respect to the state-of-the-art EE and fairness maximization techniques of \cite{EE_opt3} and \cite{fairness_3}.

 
 \textit{Notations:} Bold lowercase and uppercase letters denote vectors and matrices, respectively. $|\cdot|$ and $\text{arg}(\cdot)$ are the magnitude and the angle of a complex number, respectively (for a complex vector, they are assumed to operate element-wise). $||\mathbf{\cdot} ||_2$ is the Euclidean norm. $\mathbf{(\cdot)}^\text{T}$ is matrix transpose. $\mathbf{I}_a$ is an $a \times a$ identity matrix. $\boldsymbol{0}_{a \times b}$ and $\boldsymbol{1}_{a \times b}$ are $a \times b$ matrices all of whose elements are equal to zero and one, respectively. $\text{diag}(a_1,\dots,a_N)$ is a diagonal matrix with the elements $a_1,\dots,a_N$ on the main diagonal and zeros elsewhere. $\mathbb{C}$ and $\mathbb{R}^+$ indicate the set of the complex and non-negative real numbers, respectively. $j = \sqrt{-1}$ is the imaginary unit. $\mathbb{E}[\cdot]$ stands for the expectation operator. $\mathcal{CN}(0,b)$ represents a complex Gaussian random distribution with a mean of $0$ and a variance of $b$, and $\mathcal{U}(a,b)$ represents a uniform distribution between $a$ and $b$. $\nabla_{\mathbf a} f(\cdot)$ represents the gradient of the real-valued function $f(\cdot)$ with respect to $\mathbf a$.

\section{System Model} \label{sec:sys_model}

\begin{figure}
         \centering
         \includegraphics[width=.75\columnwidth]{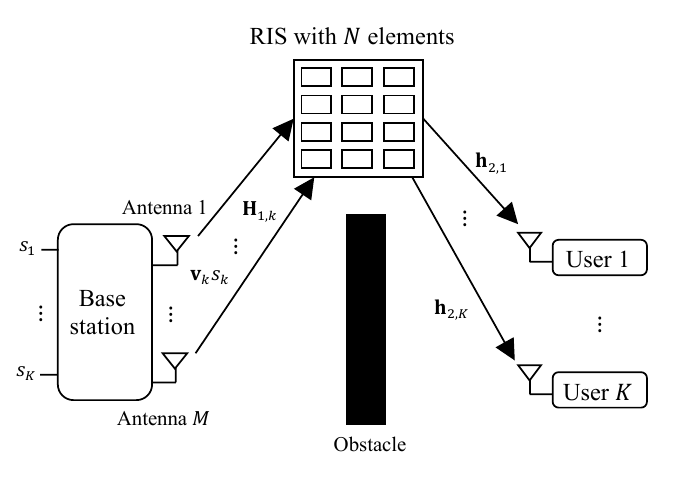}
        \caption{System model for the proposed RIS-assisted MU-MIMO mmWave communication system.}
        \label{fig:sys_model}
        \vspace{-0.2cm}
\end{figure}

We consider a downlink mmWave system in which a base station (BS) equipped with $M$ antennas communicates with $K$ single-antenna users as shown in Fig.~\ref{fig:sys_model}. The direct links between the BS and the users are assumed to be blocked and the communication is achieved via an RIS with $N$ reflecting elements. In the following subsections, we describe the received signal model and channel model.

We assume that the users are served using orthogonal resources (e.g., frequency-division multiple access (FDMA))\footnote{Orthogonal signaling limits the rate of the users, but this assumption allows us to keep the rate expressions tractable. Some other works proposed mitigating the multi-user interference using zero-forcing beamforming (ZFBF). However, the performance of ZFBF is not optimal if some RIS-user channels in the system are highly correlated, which is typical in mmWave systems. ZFBF also requires an additional hardware complexity compared to the constant-modulus analog beamformer. Interested readers can refer to \cite{mmwave_comm_survey} and \cite{AoSA_motivation} for more details.}. The signal received by user $k$ is
\begin{equation}
    y_k =  \mathbf{h}_{2,k} \boldsymbol{\Theta} \mathbf{H}_{1,k} \mathbf{v}_k \sqrt{p_k} s_k + n_k, 
    \label{eq:rec_signal}
\end{equation}
where $ \mathbf{h}_{2,k}\in \mathbb{C}^{1 \times N}$ is the channel between the $k$-th user and the RIS, $\boldsymbol{\Theta} = \text{diag}(\vartheta_1,\dots,\vartheta_N)  \in \mathbb{C}^{N \times N}$ is the RIS phase shift matrix with $\vartheta_n = e^{j \theta_n}$ and $\theta_n \in [0,2 \pi)$, $\mathbf{H}_{1,k} \in \mathbb{C}^{N \times M}$ is the BS-RIS channel corresponding to the $k$-th user, $\mathbf{v}_k  \in \mathbb{C}^{M \times 1}$ is the analog precoder for user $k$ with each element having a magnitude of $1/\sqrt{M}$, $p_k \in \mathbb{R}^+$ is the power allocated to user $k$ with $\sum_{k=1}^K p_k \leq P_\text{max}$ where $P_\text{max}$ is the total power budget, $s_k$ is the symbol to be transmitted to the $k$-th user satisfying $\mathbb{E}[|s_k|^2] = 1$, and $n_k\sim   \mathcal{C} \mathcal{N} (0,\sigma^2)$ is the complex additive white Gaussian  noise (AWGN). 

\section{Problem Formulation} \label{sec:prob_form}
In this section, we formulate the multi-objective EE and fairness optimization problem. We assume that channel state information is available at the transmitter. The rate of the $k$-th user can be expressed as
\begin{equation}
    R_k  (\mathbf{p} , \boldsymbol{\Theta}, \mathbf{v}_k) = B \log_2\left(1+\frac{1}{\sigma^2} p_k |\mathbf{h}_{2,k} \boldsymbol{\Theta} \mathbf{H}_{1,k} \mathbf{v}_k |^2\right),
    \label{rate_k}
\end{equation}
where $B$ is the bandwidth and $\mathbf{p} = [p_1,\dots,p_k]^\text{T}$.

On the other hand, the total power consumption of the system can be written as \cite{EE_opt3}
\begin{equation}
    \mathcal{P}_{\text{tot}} (\mathbf{p})  = P_\text{BS}+ \xi \sum \nolimits_{k=1}^K  p_k+ N P_{\theta}  + \sum \nolimits_{k=1}^K P_{\text{U},k} ,
    \label{poe_cons}
\end{equation}
where $P_\text{BS}$ is the total hardware static power consumption at the BS, $\xi$ is the power amplifier efficiency at the BS, $P_{\theta}$ is the power consumption of each phase shifter at the RIS, and $P_{\text{U},k}$ is the hardware static power dissipated by user $k$.

The energy efficiency (EE) is then defined as
\begin{equation}
    \eta_{\text{EE}} (\mathbf{p} , \boldsymbol{\Theta}, \mathbf{V})  = \frac{R_\text{sum}  (\mathbf{p} , \boldsymbol{\Theta}, \mathbf{V}) }{\mathcal{P}_{\text{tot}} (\mathbf{p})},
    \label{eq:EE}
\end{equation}
where $R_\text{sum}  (\mathbf{p} , \boldsymbol{\Theta}, \mathbf{V})  = \sum_{k=1}^K R_k  (\mathbf{p} , \boldsymbol{\Theta}, \mathbf{v}_k)$, and $\mathbf{V} = [\mathbf{v}_1,\dots,\mathbf{v}_k]$.

Because of the high path loss in mmWave systems, some users may have significantly lower channel gains than others. For such a situation, maximizing (\ref{eq:EE}) leads to allocating most of the resources in favor of the dominant users and ignoring those with small channel gains. As a result, achieving fairness among users in the system is also of high importance. This can be done by maximizing the following objective function:
\begin{equation}
    F (\mathbf{p} , \boldsymbol{\Theta},\mathbf{V}) = \min_{k \in \{ 1,\dots ,K \}} \frac{R_k (\mathbf{p} , \boldsymbol{\Theta}, \mathbf{v}_k) }{w_k},
    \label{eq:fairness}
\end{equation}
where $w_k$ is a \textit{weight} assigned to the $k$-th user by the service provider and represents the relative QoS of that user.

The general EE and fairness optimization problem can then be formulated as a multi-objective optimization problem as follows:
\begin{subequations}
\begin{align}
          & \max_{  \mathbf{p}, \boldsymbol{\Theta},\mathbf{V} }  \ \ \  [\eta_\text{EE} (\mathbf{p} , \boldsymbol{\Theta},\mathbf{V}), F(\mathbf{p} , \boldsymbol{\Theta},\mathbf{V}) ]  \\
        \text{s.t.} \ \ \
        & | \vartheta_n | = 1, \ \ \forall n=1,\dots,N, \label{ps_con_EEf}\\
        & \sum \nolimits_{k=1}^K p_k \leq P_\text{max}, \label{power_con_EEf}\\
        & p_k \geq 0, \ \ \forall k=1,\dots,K, \label{power_pos_EEf} \\
        & | v_{k,m} | = \frac{1}{\sqrt{M}},   \forall k=1,\dots,K,  \forall m=1,\dots,M \label{CM},%
\end{align} %
\label{all_optm_EEF} %
\end{subequations} %
 \hspace{-0.25cm}  where $ v_{k,m}$ is the $m$-th element of $\mathbf{v}_k$. This multi-objective optimization problem is challenging due to the presence of two objective functions that can have positive or negative correlation trends. To address this issue, we adopt a lexicographic approach \cite{mop}, which divides (\ref{all_optm_EEF}) into two stages. 

In the first stage, EE maximization is targeted by solving the following optimization problem:
\begin{align}
	\eta_\text{EE}^* = \max_{  \mathbf{p}, \boldsymbol{\Theta}, \mathbf{V} }  \eta_\text{EE} (\mathbf{p}, \boldsymbol{\Theta}, \mathbf{V} ) \ \text{s.t.} \ (\ref{ps_con_EEf}), (\ref{power_con_EEf}), (\ref{power_pos_EEf}), (\ref{CM}). \label{optm_EE}
\end{align}
Once $\eta_\text{EE}^*$ is obtained, the fairness can be maximized subject to an EE constraint in the second stage via 
\begin{subequations}
\begin{align}
	\max_{  \mathbf{p}, \boldsymbol{\Theta}, \mathbf{V}} &  F (\mathbf{p}, \boldsymbol{\Theta}, \mathbf{V})  \\
 \text{s.t.} \ \  & (\ref{ps_con_EEf}), (\ref{power_con_EEf}), (\ref{power_pos_EEf}), \text{and } (\ref{CM}), \\
 & \eta_\text{EE} (\mathbf{p}, \boldsymbol{\Theta}, \mathbf{V}) \geq \rho  \eta_\text{EE}^*, \label{EE_bound} %
\end{align} 
\label{optm_F} %
\end{subequations}
 \hspace{-0.33cm} where $\rho \in [0,1]$ is a design parameter. This optimization problem indicates that the fairness should be maximized at the expense of reducing the EE by a factor not greater than $(1-\rho)$.

The optimal solutions to the optimization problems~(\ref{optm_EE}) and (\ref{optm_F}) are difficult to obtain directly due to the non-convexity of these problems with respect to the optimization variables and the coupling of the optimization variables in the objective functions. In the next section, we propose an alternating optimization framework to solve (\ref{optm_EE}) and (\ref{optm_F}).

\section{Proposed Solution} \label{sec:AO}
In this section, we present the method of solving (\ref{optm_EE}) and (\ref{optm_F}) using an AO method for solving each optimization problem.

\subsection{Stage 1: EE Maximization}
\textbf{1. Power allocation:} For fixed values of the RIS phase-shift matrix and analog precoders, we target maximizing~(\ref{eq:EE}) with respect to the power allocation vector $\mathbf{p}$ subject to the constraints (\ref{power_con_EEf}) and (\ref{power_pos_EEf}). Dinkelbach's algorithm ~\cite{DB_method} can be used to efficiently solve this fractional programming problem. Specifically, the objective function (\ref{eq:EE}) can be transformed into a concave function as \footnote{In this manuscript, having some arguments omitted in a multi-variable function indicates that their values are being held fixed.}
\begin{equation}
    \mathcal{G} ( \mathbf{p},\omega^{(t-1)})  = R_\text{sum}(\mathbf{p}) - \omega^{(t-1)} \mathcal{P}_{\text{tot}}(\mathbf{p}),
    \label{obj_fun_DB_z}
\end{equation}
where $\omega^{(t)} \triangleq \eta_\text{EE}(\mathbf{p}^{(t)})$
and $\mathbf{p}^{(t)}$ denotes the value of $\mathbf{p}$ in the $t$-th iteration. The optimal power allocation can be obtained by finding the optimal value of $\mathbf{p}$ that maximizes the concave function (\ref{obj_fun_DB_z}) subject to the constraints (\ref{power_con_EEf}) and (\ref{power_pos_EEf}) (this can be achieved using  a standard algorithm, such as the interior-point method or active-set method, via the CVX toolbox \cite{cvx1}) with $ \omega^{(t)}$ updated in each iteration until convergence. A summary of the power allocation algorithm for EE maximization is given in Algorithm \ref{alg:power_allocation_EEFT}.

\begin{algorithm}[t]
\caption{Power allocation based on Dinkelbach's method for EE maximization.} \label{alg:power_allocation_EEFT}

\KwIn{ $\boldsymbol{\Theta}$, $\mathbf V$, $\mathbf p^{(1)} \!=\! \tfrac{P_{\max}}{K} \boldsymbol{1}_{K \times 1}$, $\omega^{(1)} \!=\! 0$, $\tau \!=\! 1$, $\epsilon \!>\! 0$ }

\KwOut{ $\mathbf p = \mathbf p^{(\tau)}$}

\Repeat{$|\omega^{(\tau)} - \omega^{(\tau-1)}| < \epsilon$ }{

Solve the concave optimization problem $\mathbf{p}^{(\tau+1)} =  \argmax_{ \mathbf{p} }  \mathcal{G}( \mathbf{p},\omega^{(\tau)})  \ \text{s.t.} \ (\ref{power_con_EEf}),(\ref{power_pos_EEf}) $, where $\mathcal{G} (\cdot)$ is given by~(\ref{obj_fun_DB_z})\;

$\omega^{(\tau+1)} = \eta_{\text{EE}}(\mathbf{p}^{(\tau+1)},\boldsymbol{\Theta} ,\mathbf{V})$\;

$\tau \leftarrow \tau+1$\;
}
\end{algorithm}


\textbf{2. Phase shift matrix:} Since the phase shift matrix appears only in the numerator of (\ref{eq:EE}), it is sufficient to maximize $R_\text{sum}$ subject to the constraint (\ref{ps_con_EEf}).

 We use the following substitutions for simplicity of illustration:
\begin{equation}
    \mathbf{H}_{1,k} \mathbf{v}_k =[b_{1,k}e^{j\gamma_{1,k}}, b_{2,k}e^{j\gamma_{2,k}},\dots,b_{N,k}e^{j\gamma_{N,k}}]^\text{T},
    \label{sub1}
\end{equation}
\begin{equation}
    \mathbf{h}_{2,k} = [c_{1,k}e^{j\mu_{1,k}}, c_{2,k}e^{j\mu_{2,k}},\dots,c_{N,k}e^{j\mu_{N,k}}],
    \label{sub2}
\end{equation}
where $b_{n,k}, c_{n,k} \in \mathbb{R}^+$, and $ \gamma_{n,k}, \mu_{n,k}  \in [0,2\pi)$. Using (\ref{sub1}) and (\ref{sub2}), we can express the effective channel gain as
\begin{equation}
    \begin{split}
        h_k  & =  |\mathbf{h}_{2,k} \boldsymbol{\Theta} \mathbf{H}_{1,k} \mathbf{v}_k |^2 \\
        & 
         \begin{split}
         = \sum_{i=1}^{N} b^2_{i,k}c^2_{i,k} + & 2 \sum_{m=1}^{N-1}  \sum_{\ell=m+1}^{N}   \big[ b_{m,k}c_{m,k} b_{\ell,k}c_{\ell,k} \\
         & \cos (\theta_m - \theta_\ell + \psi_{m,k} - \psi_{\ell,k}) \big],
         \end{split}
    \end{split}
    \label{y_k_def}
\end{equation}
where $\psi_{i,k} =\gamma_{i,k} + \mu_{i,k}$. Therefore
\begin{equation}
    \frac{\partial h_k}{\partial \theta_n} (\boldsymbol{\theta})  = \! -2 b_{n,k}c_{n,k} \sum_{i=1, i \neq n}^{N} \!\!\!\!  b_{i,k}c_{i,k} \sin (\theta_n - \theta_i + \psi_{n,k} - \psi_{i,k}),
\end{equation}
and
\begin{equation}
    \frac{\partial R_k}{\partial \theta_n} (\boldsymbol{\theta}) = \frac{1}{\ln 2} \bigg( \frac{p_k}{\sigma^2+ p_kh_k} \bigg) \frac{\partial h_k}{\partial \theta_n} (\boldsymbol{\theta}),
    \label{PD_Rk}
\end{equation}
where $\boldsymbol{\theta} = [\theta_1,\dots, \theta_n]^\text{T}$. Also, we can obtain
\begin{equation}
    \frac{\partial R_\text{sum}}{\partial \theta_n} (\boldsymbol{\theta}) = \sum_{k=1}^K \frac{\partial R_k}{\partial \theta_n} (\boldsymbol{\theta}),
    \label{PD_org}
\end{equation}
and the update of $\boldsymbol{\theta}$ should follow
\begin{equation}
   \boldsymbol{\theta}^{(t+1)} = \boldsymbol{\theta}^{(t)} + \alpha \nabla_{\boldsymbol{\theta}} R_\text{sum} ( \boldsymbol{\theta}^{(t)}),
    \label{update_rule}
\end{equation}
where $\boldsymbol{\theta}^{(t)}$ denotes the value of $\boldsymbol{\theta}$ at the $t$-th iteration, $\alpha>0$ is the step size, and $\nabla_{\boldsymbol{\theta}} R_\text{sum} ( \boldsymbol{\theta}) = [\frac{\partial R_\text{sum}}{\partial \theta_1} (\boldsymbol{\theta}),\dots,\frac{\partial R_\text{sum}}{\partial \theta_N} (\boldsymbol{\theta})]^\text{T}$. 

\textbf{3. Analog precoder:} The $k$-th user's analog precoder optimization problem is
\begin{align}
	\max_{ \mathbf{v}_k }  |\mathbf{h}_{2,k} \boldsymbol{\Theta} \mathbf{H}_{1,k} \mathbf{v}_k |^2 \ \text{s.t.} \ (\ref{CM} ) .   \label{APC_obj} %
\end{align}
Let
\begin{equation}
    \mathbf{h}_{2,k} \boldsymbol{\Theta} \mathbf{H}_{1,k} = [s_1e^{j\beta_1},\dots,s_Me^{j\beta_M}],
    \label{eq:ang_ali}
\end{equation}
and
\begin{equation}
    \mathbf{v}_{k} = \frac{1}{\sqrt{M}} [e^{j\varphi_1},\dots,e^{j\varphi_M}],
\end{equation}
where $ s_n \in \mathbb{R}^+$ and $  \beta_n, \varphi_n \in [0, 2 \pi)$. Then the objective function in (\ref{APC_obj}) can be written as
\begin{equation}
    |\mathbf{h}_{2,k} \boldsymbol{\Theta} \mathbf{H}_{1,k} \mathbf{v}_k |^2 \! = \! \frac{1}{\sqrt{M}} \bigg| \! \sum_{i=1}^{M}s_i e^{j(\varphi_i + \beta_i)} \! \bigg|^2 \! \leq \! \frac{1}{\sqrt{M}}\bigg|\sum_{i=1}^{M}s_i \bigg|^2 ,
\end{equation}
where the upper bound is obtained (using the triangle inequality) by setting $\varphi_m = - \beta_m \  \forall m = 1,\dots,M$.

\subsection{Stage 2: Fairness Maximization with an EE Constraint}
\textbf{1. Power allocation:} Next, we target the optimization problem (\ref{optm_F}) with respect to the power allocation vector $\mathbf{p}$. To solve this problem, we introduce an intermediate variable to represent the minimum weighted rate as
\begin{equation}
    z = \min_{k \in \{ 1,\dots ,K \} } \frac{R_k} { w_k},
    \label{ub_z}
\end{equation}
so that (\ref{optm_F}) becomes
\begin{subequations}
\begin{align}
           \max_{  \mathbf{p},z}  \ \ \ & z  \\
        \text{s.t.} \ \ \
        & (\ref{power_con_EEf}) \text{ and } (\ref{power_pos_EEf}), \\
        & R_\text{sum} (\mathbf{p}) \geq \rho  \eta_\text{EE}^*\mathcal{P}_{\text{tot}}(\mathbf{p}), \label{conn1} \\
        & \frac{1}{w_k} R_k(\mathbf{p}) \geq z, \ \ \forall k =1,\dots,K, \label{conn2} %
\end{align}
\end{subequations}
which is a concave optimization problem amenable to solution via standard solvers such as CVX.

\textbf{2. Phase shift matrix:} As the next step in the AO approach, we consider the optimization problem (\ref{optm_F}) with respect to the phase shift matrix. It is obvious that the derivative of the minimum function in (\ref{eq:fairness}) is non-smooth and has discontinuities at points where $\frac{R_i}{w_i} = \frac{R_k}{w_k}$, where $i,k \in \{ 1,\dots,K \}$ and $i \neq k$. Therefore, the gradient ascent method cannot be applied directly. To resolve this issue, we use the well-known log-sum-exp (LSE) approximation to approximate $F$ as \cite{LSE}
\begin{equation}
    F \approx \hat{F} = - \frac{1}{\zeta} \ln \left( \sum \nolimits_{k=1}^K e^{- \zeta R_k / w_k} \right),
\end{equation}
where $\zeta > 0$. This approximation is accurate for sufficiently large values of $\zeta$.

The partial derivative of $\hat{F}$ with respect to $\theta_n$ can be written as
\begin{equation}
    \frac{\partial \hat{F}}{\partial \theta_n} (\boldsymbol{\theta}) = \frac{\sum_{k=1}^K \frac{1}{w_k} \frac{\partial R_k}{\partial \theta_n}  e^{- \zeta R_k / w_k}}{\sum_{k=1}^K e^{- \zeta R_k / w_k}},
    \label{eq:drmw}
\end{equation}
which is a continuous function of $\boldsymbol{\theta}$. Therefore, the expression (\ref{eq:drmw}) can be used (with a large value of $\zeta$) as an approximation to the partial derivative $\frac{\partial F}{\partial \theta_n}$.

 The update of $\boldsymbol{\theta}$ should then follow
\begin{equation}
   \boldsymbol{\theta}^{(t+1)} = \boldsymbol{\theta}^{(t)} + \alpha \nabla_{\boldsymbol{\theta}} \hat{F}(\boldsymbol{\theta}^{(t)}),
\end{equation}
where $\nabla_{\boldsymbol{\theta}} \hat{F}(\boldsymbol{\theta}) = [ \frac{\partial \hat{F}}{\partial \theta_1} (\boldsymbol{\theta}),\dots, \frac{\partial \hat{F}}{\partial \theta_N} (\boldsymbol{\theta})]^\text{T}$. However, the constraint (\ref{EE_bound}) should be checked after each update of $\boldsymbol{\theta}$; if it is violated by the new value of $\boldsymbol{\theta}$, the update is not accepted by the algorithm.

\textbf{3. Analog precoder:} Since it is required to maximize the effective channel gain of the user with the minimum weighted rate, we can follow the same procedure as in EE maximization (Staege 1) to obtain the analog precoder.  

The two-stage optimization procedure is summarized in Algorithm~\ref{alg:AO}.

\begin{algorithm}[t]
\caption{Lexicographic approach for EE and fairness maximization.} \label{alg:AO}

\KwIn{ $\zeta \geq 0$, $\rho \in [0,1]$, $\alpha > 0$, $\boldsymbol{\Theta}^{(1)} = \mathbf I_N$, $\mathbf V^{(1)} = \tfrac{1}{\sqrt{M}} \boldsymbol{1}_{M \times K}$, $\eta_{\mathrm{EE}}^{(1)} = 0$, $t = 1$ $\epsilon > 0$ }

\KwOut{ $\mathbf p = \mathbf p^{(t)}$, $\boldsymbol{\Theta} = \boldsymbol{\Theta}^{(t)}$, $\mathbf V = \mathbf V^{(t)}$}

\tcc{Stage 1: EE maximization}

\Repeat{$|\eta_\text{EE}^{(t)} - \eta_\text{EE}^{(t-1)}| < \epsilon$ }{

	Compute $\mathbf p^{(t+1)}$ using Algorithm~\ref{alg:power_allocation_EEFT}\;

	$\boldsymbol{\theta}^{(1)} \leftarrow \boldsymbol{0}_{N \times 1}$, $i = 1$\;

	\Repeat{$\| \boldsymbol{\theta}^{(i)} - \boldsymbol{\theta}^{(i-1)} \|_2 < \epsilon$}{
	
		$\boldsymbol{\theta}^{(i+1)} = \boldsymbol{\theta}^{(i)} + \alpha \nabla_{\boldsymbol{\theta}}R_{\mathrm{sum}}\big(\boldsymbol{\theta}^{(i)} \big)$	\;
		
		$i \leftarrow i+1$\;
	
	}	
	
	$\boldsymbol{\Theta}^{(t+1)} = \mathrm{diag}\big(\exp \big(j\boldsymbol{\theta}^{(i)} \big)\big)$\;
	
	\For{$k = 1, 2, \ldots, K$}{
	
		$\boldsymbol{\beta}_k = \text{arg}(\mathbf{h}_{2,k} \boldsymbol{\Theta} \mathbf{H}_{1,k})$\;
		
		$\mathbf{v}_k^{(t+1)} = \frac{1}{\sqrt{M}} e^{-j\boldsymbol{\beta}_k}$\;
    	
    }
    
    $\mathbf{V}^{(t+1)} = [\mathbf{v}_1^{(t+1)},\dots,\mathbf{v}_K^{(t+1)}]$\;
    
    Compute $\eta_\text{EE}^{(t+1)} (  \mathbf{p}^{(t+1)} , \!\boldsymbol{\Theta}^{(t+1)},\!\mathbf{V}^{(t+1)})$ using~(\ref{eq:EE})\;

	$t \leftarrow t + 1$\;
	
}

$\eta_\text{EE}^* =  \eta_\text{EE}^{(t)}$\;

\tcc{Stage 2: Fairness maximization with EE constraint}

$F^{(t)} = F (\mathbf{p}^{(t)} , \boldsymbol{\Theta}^{(t)},\mathbf{V}^{(t})$\;

\Repeat{$|F^{(t)} - F^{(t-1)}|< \epsilon$}{

	Solve the concave optimization problem $\big( \mathbf p^{(t+1)},  z^{(t+1)}\big) = \underset{\mathbf p, z}{\argmax} \ \ z \ \  \text{s.t.} \ (\ref{power_con_EEf}), \! (\ref{power_pos_EEf}), \! (\ref{conn1}), \! (\ref{conn2}) $;
	
	$\boldsymbol{\theta}^{(1)} =  \boldsymbol{0}_{N \times 1}$\;
	
	\Repeat{$\| \boldsymbol{\theta}^{(i)} - \boldsymbol{\theta}^{(i-1)}\|_2 < \epsilon$}{
	
		$ \boldsymbol{\theta}^{(i+1)} =  \boldsymbol{\theta}^{(i)} + \alpha \nabla_{\boldsymbol{\theta}} \hat{F} \big(\boldsymbol{\theta}^{(i)}\big)$\;
		
		Compute $ \eta_\text{EE} (\boldsymbol{\theta}^{(i+1)})$ using~(\ref{eq:EE})\;
		
		\If{$ \eta_\text{EE} (\boldsymbol{\theta}^{(i+1)}) \geq \rho \eta_\text{EE}^*$}{
    			$ \boldsymbol{\theta}^{(i+1)} =  \boldsymbol{\theta}^{(i)} - \alpha \nabla_{\boldsymbol{\theta}} \hat{F} (\boldsymbol{\theta}^{(i)})$\;
    			
    			\Break\;
  		}
  		
  		$i \leftarrow i+1$\;
	}
	
	$\boldsymbol{\Theta}^{(t+1)} = \text{diag} (e^{ j\boldsymbol{\theta}^{(i)}})$\;
	
	Compute $\mathbf{V}^{(t+1)}$ as per lines 9-13\;
	
	Compute~$F^{(t+1)} (  \! \mathbf{p}^{(t+1)} , \! \boldsymbol{\Theta}^{(t+1)}, \! \boldsymbol{V}^{(t+1)}\!)$~using~(\ref{eq:fairness})\;

	$t \leftarrow t+1$\;

}

\end{algorithm}

\begin{figure*}[t]
\begin{minipage}[t]{0.32\textwidth}%
\begin{center}
\newcommand{\vasymptote}[2][]{
    \draw [densely dashed,#1] ({rel axis cs:0,0} -| {axis cs:#2,0}) -- ({rel axis cs:0,1} -| {axis cs:#2,0});
}
\resizebox{0.99\columnwidth}{!}{
\begin{tikzpicture}
  \begin{axis}[
    xlabel={Iteration number},
	xtick={0,3,6,9,12,15},
	xmin=0,xmax=15,
	ylabel={Average EE [Mbits/s/Joule]},
	ylabel near ticks,
	ytick={0,40,80,120},
	ymin=0,ymax=120,
	grid=both,
	minor grid style={gray!25},
	major grid style={gray!25},
	legend columns=1, 
legend style={{nodes={scale=0.9, transform shape}}, at={(0.07,0)},  anchor=south west, draw=black,fill=white,legend cell align=left,inner sep=1pt,row sep = -2pt}]
	\addplot[line width=1pt,solid,color=black,mark=o,mark options={solid,black},mark size=2pt] table [y=maxEE, x=iter,col sep = comma]{conv_EE.csv};
	\addlegendentry{EE maximization (Stage 1)}
	\addplot[line width=1pt,solid,color=red,mark=triangle,mark options={solid,red},mark size=2.5pt] table [y=maxF85, x=iter,col sep = comma]{conv_EE.csv};
	\addlegendentry{Fairness maximization (Stage 2), $\rho=0.85$}
	\addplot[line width=1pt,solid,color=blue,mark=square,mark options={solid,blue},mark size=2pt] table [y=maxF50, x=iter,col sep = comma]{conv_EE.csv};
	\addlegendentry{Fairness maximization (Stage 2), $\rho=0.5$}
	\end{axis}
\end{tikzpicture}
}\vspace{-0.6cm}
\par\end{center}
\caption{Average EE vs. iteration number for EE maximization (Stage~1) and fairness maximization (Stage~2) at $P_{\max} = 25$~dBm.}
\label{fig:convEE}%
\end{minipage}\hfill{}%
\begin{minipage}[t]{0.32\textwidth}%
\begin{center}
\newcommand{\vasymptote}[2][]{
    \draw [densely dashed,#1] ({rel axis cs:0,0} -| {axis cs:#2,0}) -- ({rel axis cs:0,1} -| {axis cs:#2,0});
}
\resizebox{0.99\columnwidth}{!}{
\begin{tikzpicture}
  \begin{axis}[
    xlabel={Iteration number},
	xtick={0,3,6,9,12,15},
	xmin=0,xmax=15,
	ylabel={Avg. min. weighted rate [Mbits/s]},
	ylabel near ticks,
	ytick={0,5,10,15,20,26},
	ymin=0,ymax=26,
	grid=both,
	minor grid style={gray!25},
	major grid style={gray!25},
	legend columns=1, 
legend style={{nodes={scale=0.9, transform shape}}, at={(0.07,0.1)},  anchor=south west, draw=black,fill=white,legend cell align=left,inner sep=1pt,row sep = -2pt}]
	\addplot[line width=1pt,solid,color=blue,mark=square,mark options={solid,blue},mark size=2pt] table [y=maxF50, x=iter,col sep = comma]{conv_mwr.csv};
	\addlegendentry{Fairness maximization (Stage 2), $\rho=0.5$}
	\addplot[line width=1pt,solid,color=red,mark=triangle,mark options={solid,red},mark size=2.5pt] table [y=maxF85, x=iter,col sep = comma]{conv_mwr.csv};
	\addlegendentry{Fairness maximization (Stage 2), $\rho=0.85$}
	\addplot[line width=1pt,solid,color=black,mark=o,mark options={solid,black},mark size=2pt] table [y=maxEE, x=iter,col sep = comma]{conv_mwr.csv};
	\addlegendentry{EE maximization (Stage 1)}
	\end{axis}
\end{tikzpicture}
}\vspace{-0.6cm}
\par\end{center}
\caption{Avg. min. weighted rate vs. iteration number for EE maximization (Stage~1) and fairness maximization (Stage~2) at $P_{\max} = 25$~dBm.}
\label{fig:convMWR}%
\end{minipage}\hfill{}%
\begin{minipage}[t]{0.32\textwidth}%
\begin{center}
\newcommand{\vasymptote}[2][]{
    \draw [densely dashed,#1] ({rel axis cs:0,0} -| {axis cs:#2,0}) -- ({rel axis cs:0,1} -| {axis cs:#2,0});
}
\resizebox{0.99\columnwidth}{!}{
\begin{tikzpicture}
  \begin{axis}[
    xlabel={$P_{\max}$~[dBm]},
	xtick={5,15,25,35,45},
	xmin=5,xmax=45,
	ylabel={Average EE [Mbits/s/Joule]},
	ylabel near ticks,
	ytick={0,40,80,120,160,200},
	ymin=0,ymax=200,
	grid=both,
	minor grid style={gray!25},
	major grid style={gray!25},
	legend columns=1, 
legend style={{nodes={scale=0.9, transform shape}}, at={(0,0.749)},  anchor=south west, draw=black,fill=white,legend cell align=left,inner sep=1pt,row sep = -2pt}]
	\addplot[line width=1pt,solid,color=black,mark=o,mark options={solid,black},mark size=2pt] table [y=maxEE, x=P,col sep = comma]{P_EE.csv};
	\addlegendentry{EE maximization~\cite{EE_opt3}}
	\addplot[line width=1pt,solid,color=red,mark=triangle,mark options={solid,red},mark size=2.5pt] table [y=prop85, x=P,col sep = comma]{P_EE.csv};
	\addlegendentry{Proposed, $\rho=0.85$}
	\addplot[line width=1pt,solid,color=blue,mark=square,mark options={solid,blue},mark size=2pt] table [y=prop50, x=P,col sep = comma]{P_EE.csv};
	\addlegendentry{Proposed, $\rho=0.5$}
	\addplot[line width=1pt,solid,color=mygreen,mark=diamond,mark options={solid,mygreen},mark size=2.5pt] table [y=eeAware, x=P,col sep = comma]{P_EE.csv};
	\addlegendentry{EE-aware fairness maximization~~\cite{fairness_3}}		
	\end{axis}
\end{tikzpicture}
}\vspace{-0.6cm}
\par\end{center}
\caption{Average EE as the power budget varies for EE maximization~\cite{EE_opt3}, EE-aware fairness maximization~\cite{fairness_3}, and the proposed approach.}
\label{fig:EEvsP}%
\end{minipage}
\end{figure*}

\begin{figure*}[t]
\begin{minipage}[t]{0.32\textwidth}%
\begin{center}
\newcommand{\vasymptote}[2][]{
    \draw [densely dashed,#1] ({rel axis cs:0,0} -| {axis cs:#2,0}) -- ({rel axis cs:0,1} -| {axis cs:#2,0});
}
\resizebox{0.99\columnwidth}{!}{
\begin{tikzpicture}
  \begin{axis}[
    xlabel={$P_{\max}$~[dBm]},
	xtick={5,15,25,35,45},
	xmin=5,xmax=45,
	ylabel={Average Jain's fairness index},
	ylabel near ticks,
	ytick={0,0.2,0.4,0.6,0.8,1},
	ymin=0,ymax=1,
	grid=both,
	minor grid style={gray!25},
	major grid style={gray!25},
	legend columns=1, 
legend style={{nodes={scale=0.9, transform shape}}, at={(0.158,0)},  anchor=south west, draw=black,fill=white,legend cell align=left,inner sep=1pt,row sep = -2pt}]
	\addplot[line width=1pt,solid,color=mygreen,mark=diamond,mark options={solid,mygreen},mark size=2.5pt] table [y=eeAware, x=P,col sep = comma]{P_JFI.csv};
	\addlegendentry{EE-aware fairness maximization~~\cite{fairness_3}}
	\addplot[line width=1pt,solid,color=blue,mark=square,mark options={solid,blue},mark size=2pt] table [y=prop50, x=P,col sep = comma]{P_JFI.csv};
	\addlegendentry{Proposed, $\rho=0.5$}
	\addplot[line width=1pt,solid,color=red,mark=triangle,mark options={solid,red},mark size=2.5pt] table [y=prop85, x=P,col sep = comma]{P_JFI.csv};
	\addlegendentry{Proposed, $\rho=0.85$}
	\addplot[line width=1pt,solid,color=black,mark=o,mark options={solid,black},mark size=2pt] table [y=maxEE, x=P,col sep = comma]{P_JFI.csv};
	\addlegendentry{EE maximization~\cite{EE_opt3}}		
	\end{axis}
\end{tikzpicture}
}\vspace{-0.6cm}
\par\end{center}
\caption{Average Jain's fairness index as the power budget varies for EE maximization~\cite{EE_opt3}, EE-aware fairness maximization~\cite{fairness_3}, and the proposed approach.}
\label{fig:JFIvsP}%
\end{minipage}\hfill{}%
\begin{minipage}[t]{0.32\textwidth}%
\begin{center}
\newcommand{\vasymptote}[2][]{
    \draw [densely dashed,#1] ({rel axis cs:0,0} -| {axis cs:#2,0}) -- ({rel axis cs:0,1} -| {axis cs:#2,0});
}
\resizebox{0.99\columnwidth}{!}{
\begin{tikzpicture}
  \begin{axis}[
    xlabel={Number of RIS elements},
	xtick={4,100,200,300,400},
	xmin=4,xmax=400,
	ylabel={Average EE [Mbits/s/Joule]},
	ylabel near ticks,
	ytick={0,100,200,300,400,450},
	ymin=0,ymax=450,
	grid=both,
	minor grid style={gray!25},
	major grid style={gray!25},
	legend columns=1, 
legend style={{nodes={scale=0.9, transform shape}}, at={(0.175,0)},  anchor=south west, draw=black,fill=white,legend cell align=left,inner sep=1pt,row sep = -2pt}]
	\addplot[line width=1pt,solid,color=black,mark=o,mark options={solid,black},mark size=2pt] table [y=maxEE, x=nRIS,col sep = comma]{N_EE.csv};
	\addlegendentry{EE maximization~\cite{EE_opt3}}
	\addplot[line width=1pt,solid,color=red,mark=triangle,mark options={solid,red},mark size=2.5pt] table [y=prop85, x=nRIS,col sep = comma]{N_EE.csv};
	\addlegendentry{Proposed, $\rho=0.85$}
	\addplot[line width=1pt,solid,color=blue,mark=square,mark options={solid,blue},mark size=2pt] table [y=prop50, x=nRIS,col sep = comma]{N_EE.csv};
	\addlegendentry{Proposed, $\rho=0.5$}
	\addplot[line width=1pt,solid,color=mygreen,mark=diamond,mark options={solid,mygreen},mark size=2.5pt] table [y=eeAware, x=nRIS,col sep = comma]{N_EE.csv};
	\addlegendentry{EE-aware fairness maximization~\cite{fairness_3}}
	\end{axis}
\end{tikzpicture}
}\vspace{-0.6cm}
\par\end{center}
\caption{Average EE as the number of RIS elements increases for EE maximization~\cite{EE_opt3}, EE-aware fairness maximization~\cite{fairness_3}, and the proposed approach.}
\label{fig:EEvsN}%
\end{minipage}\hfill{}%
\begin{minipage}[t]{0.32\textwidth}%
\begin{center}
\newcommand{\vasymptote}[2][]{
    \draw [densely dashed,#1] ({rel axis cs:0,0} -| {axis cs:#2,0}) -- ({rel axis cs:0,1} -| {axis cs:#2,0});
}
\resizebox{0.99\columnwidth}{!}{
\begin{tikzpicture}
  \begin{axis}[
    xlabel={Number of RIS elements},
	xtick={4,100,200,300,400},
	xmin=4,xmax=400,
	ylabel={Average Jain's fairness index},
	ylabel near ticks,
	ytick={0.2,0.4,0.6,0.8,1},
	ymin=0.2,ymax=1,
	grid=both,
	minor grid style={gray!25},
	major grid style={gray!25},
	legend columns=1, 
legend style={{nodes={scale=0.9, transform shape}}, at={(0.158,0)},  anchor=south west, draw=black,fill=white,legend cell align=left,inner sep=1pt,row sep = -2pt}]
	\addplot[line width=1pt,solid,color=mygreen,mark=diamond,mark options={solid,mygreen},mark size=2.5pt] table [y=eeAware, x=nRIS,col sep = comma]{N_JFI.csv};
	\addlegendentry{EE-aware fairness maximization~~\cite{fairness_3}}
	\addplot[line width=1pt,solid,color=blue,mark=square,mark options={solid,blue},mark size=2pt] table [y=prop50, x=nRIS,col sep = comma]{N_JFI.csv};
	\addlegendentry{Proposed, $\rho=0.5$}	
	\addplot[line width=1pt,solid,color=red,mark=triangle,mark options={solid,red},mark size=2.5pt] table [y=prop85, x=nRIS,col sep = comma]{N_JFI.csv};
	\addlegendentry{Proposed, $\rho=0.85$}	
	\addplot[line width=1pt,solid,color=black,mark=o,mark options={solid,black},mark size=2pt] table [y=maxEE, x=nRIS,col sep = comma]{N_JFI.csv};
	\addlegendentry{EE maximization~\cite{EE_opt3}}
	\end{axis}
\end{tikzpicture}
}\vspace{-0.6cm}
\par\end{center}
\caption{Average Jain's fairness index as the number of RIS elements increases for EE maximization~\cite{EE_opt3}, EE-aware fairness maximization~\cite{fairness_3}, and the proposed approach.}
\label{fig:JFIvsN}%
\end{minipage}
\end{figure*}

\section{Numerical Results} \label{sec:sim}
In this section, the performance of the proposed method is demonstrated and compared with that of two benchmark schemes, the pure EE maximization scheme of \cite{EE_opt3}, and the EE-aware fairness maximization scheme of \cite{fairness_3}.

A square cell with a side of \unit[150]{m} is considered in simulations with a BS located at the point $(0,0,10)$ m having $M=16$ antennas that serve four users with $w_k \sim \mathcal{U}(1,4)$. FDMA transmission is used with a center frequency for the $k$-th user of $f_k = \unit[27.75 + (2k-1) 0.0625]{GHz}$ (each user has a bandwidth of \unit[125]{MHz}). A spacing of $\mathbf{\lambda_c}/2$ is assumed between two adjacent antennas at the BS, where $\lambda_c $ is the carrier wavelength for the system center frequency $f_c = \unit[28]{GHz}$. The number of RIS elements is assumed to be $N=64$ with both vertical and horizontal spacings of $\mathbf{\lambda_c}/2$ between adjacent elements. We assume that the locations of the users are  uniformly distributed  in the cuboidal region with opposite corners at points $(30,-75,0)$ m and $(150,75,2)$ m. The total hardware static power at the BS is $P_\text{BS} = \unit[9]{dBW}$, while that of the $k$-th UE is $P_{\text{U},k} = \unit[10]{dBm}$. The power consumption of each RIS phase shifter is assumed to be $P_\theta  = \unit[1]{dBm}$ and the circuit dissipated power coefficient at the BS $\xi$ is set to $1.2$ \cite{EE_opt3}. The minimum function approximation parameter $\zeta$ is set to $50$ and the algorithmic convergence threshold is selected to be $\epsilon = 1 \times 10^{-3}$. 

Monte Carlo simulation is performed to obtain the results by averaging the performance metric over 1000 combinations of random channels and user locations. The widely used Saleh-Valenzuela (SV) channel model \cite{ch_model_2} is adopted, where the number of paths is assumed to be $L = 4$ \cite{EE_opt3,EE_opt4}. The path gain of the LoS component is calculated using the free space path loss formula, and a difference of \unit[15]{dBm} is assumed between the LoS and NLoS path gains \cite{LOS_NLOS}. The small-scale fading is modeled as a Gaussian random variable with a \unit[0]{dB} mean and a \unit[1]{dB} variance.

Two metrics are used to compare the performance of the different methods; the EE (defined in (\ref{eq:EE})) and Jain's fairness index for weighted rates, which is defined as \cite{sim5}
\begin{equation}
    F_J = \frac{(\sum_{k=1}^K R_k / w_k)^2}{K \sum_{k=1}^K (R_k / w_k)^2} \in [1/K,1],
\end{equation}
where $F_J = 1$ indicates that all users have the same weighted rate $R_k / w_k$.

To examine the convergence of the objective functions (\ref{eq:EE}) and (\ref{eq:fairness}) during Algorithm \ref{alg:AO}, we plot the EE and the minimum weighted rate versus the iteration number in Figs.~\ref{fig:convEE} and~\ref{fig:convMWR}, respectively, for the EE maximization problem (Stage 1) and the fairness maximization problem (Stage 2) for two values of $\rho$ ($0.85$ and $0.5$) and $P_\text{max} = \unit[25]{dBm}$. It can be noted that in all cases the proposed algorithm converges within a small number of iterations. In addition, the minimum weighted rate produced by allocating the resources to maximize the EE (Stage 1) is very low. However, when 15\% and 50\% maximum reductions in the EE are allowed (corresponding to $\rho=0.85$ and $\rho = 0.5$, respectively), the minimum weighted rate can be increased significantly. The actual reductions in the EE are $13$\% and $42$\%  for $\rho = 0.85$ and $\rho = 0.5$ respectively, which are less than maximum allowed reductions $1-\rho$.

%
%
%
%
%

Figs.~\ref{fig:EEvsP} and~\ref{fig:JFIvsP} show the EE and Jain's fairness index of the proposed method versus the maximum transmit power, respectively, for $\rho = 0.85$ and $\rho = 0.5$. The proposed method is compared with the benchmark algorithms \cite{EE_opt3} and \cite{fairness_3}. It can be noticed that all curves saturate when $P_\text{max}$ exceeds \unit[40]{dBm} as the additional power does not help in increasing the EE objective function (\ref{eq:EE}). On the one hand, the method in \cite{EE_opt3} achieves higher EE than the other algorithms. However, it produces a very low fairness index, especially at low power budgets. On the other hand, \cite{fairness_3} achieves excellent fairness even at low power budgets, but lower EE than the other methods. The proposed method achieves a controllable trade-off between the aforementioned extreme situations. If the value of $\rho$ is chosen carefully (e.g., $\rho = 0.85$), a very good level of fairness can be attained with with a pre-designed minor reduction in the EE. 

%

Figs.~\ref{fig:EEvsN} and~\ref{fig:JFIvsN} depict the EE and Jain's fairness index for $P_\text{max} = \unit[25]{dBm}$ as a function of the number of RIS elements, respectively. The relationship between the EE and the number of RIS elements follows a log-like trend. Increasing the number of RIS elements makes the numerator of (\ref{eq:EE}) increase logarithmically while the denominator increases linearly. Since the power consumption of the RIS elements is small, the ratio (\ref{eq:EE}) saturates. In other terms, adding extra RIS elements can have a significant contribution towards boosting both the EE and fairness. In the case of fairness maximization, a small number of RIS elements (i.e., $N \leq 50$) is sufficient to achieve a Jain's fairness score of 1. A larger number of of RIS elements allows the proposed method to achieve a very good level of fairness (i.e., higher than 0.9) with higher fairness for $\rho = 0.5$ than for $\rho = 0.85$. This comes at the expense of a minor penalty in the EE as can be seen in  Fig.~\ref{fig:EEvsN}.


\balance

\section{Conclusion} \label{sec:conc}
This paper presents a two-stage lexicographic approach to find the optimal power allocation, RIS phase shift matrix, and analog precoders that maximize both EE and fairness in an RIS-assisted mmWave MU-MISO system. First, the EE is maximized, then the minimum weighted rate is maximized subject to a constraint on the allowable reduction in the EE. The optimization problems in both stages are solved via an AO procedure that uses Dinkelbach's method for power allocation, gradient ascent for RIS phase shift matrix optimization, and beam alignment for analog precoder design. The results show that the proposed method can achieve a better trade-off between EE and fairness compared to methods that optimize only one of these metrics.

\section*{Acknowledgement}
The authors would like to thank Dr. Arman Farhang for helpful discussions.

\bibliographystyle{IEEEtran}
\bibliography{Bibliography}
\end{document}